# The Origin of Cosmic Rays: What Can GLAST Say?

Jonathan F. Ormes, Seth Digel, Igor V. Moskalenko[1], and Alexander Moiseev,
*Goddard Space Flight Center, Greenbelt, Maryland*
and Roger Williamson
*Stanford University*
on behalf of the GLAST collaboration

**Abstract.** Gamma rays in the band from 30 MeV to 300 GeV, used in combination with direct measurements and with data from radio and X-ray bands, provide a powerful tool for studying the origin of Galactic cosmic rays. Gamma-ray Large Area Space Telescope (GLAST) with its fine 10-20 arcmin angular resolution will be able to map the sites of acceleration of cosmic rays and their interactions with interstellar matter. It will provide information that is necessary to study the acceleration of energetic particles in supernova shocks, their transport in the interstellar medium and penetration into molecular clouds.

## INTRODUCTION

The identification of the sites of cosmic-ray acceleration is one of the main unsolved problems in Galactic cosmic-ray astrophysics. We know from radio [1] and recent X-ray observations [2] of synchrotron radiation from supernova remnants that electrons are accelerated to TeV energies. Direct evidence for the acceleration of protons is not yet in hand. There is a considerable theoretical literature that quantifies how this acceleration takes place in the turbulent magnetic fields associated with the shock waves generated by supernova explosions propagating into the interstellar medium [3]. With the Gamma-ray Large Area Space Telescope (GLAST), proposed for launch in 2005, we have the possibility of detecting γ rays from the freshly accelerated cosmic-ray nuclei at their acceleration site. The optimal candidate sites will be those where the shock waves collide with either swept up interstellar matter or nearby clouds. The most commonly discussed site is individual supernova remnants [4] however multiple supernovas inside superbubbles have also been considered [5].

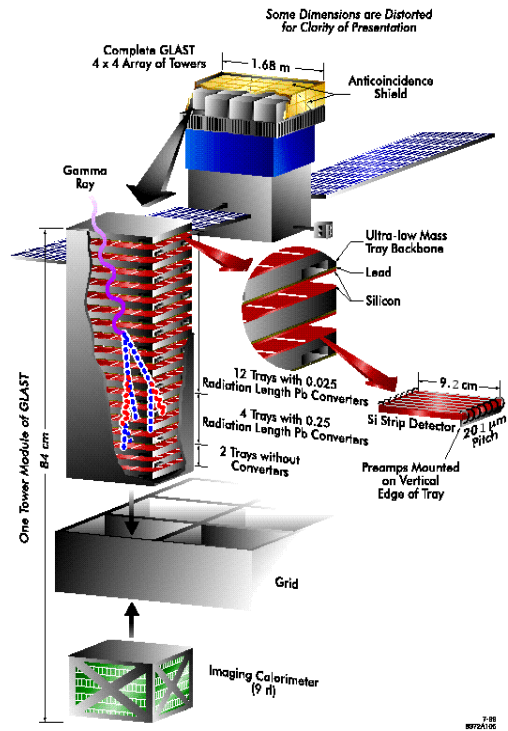

Fig. 1: The GLAST large area telescope

---

[1] NRC Resident Research Associate visiting GSFC from Inst. for Nucl. Phys., Moscow State U.

While the great majority of the cosmic rays are nuclei, not electrons, most of the electromagnetic signatures of cosmic rays reflect electrons as seen in synchrotron radiation or inverse Compton (IC) interactions of photons with electrons. This includes even the highest energy band, photons around 1 TeV. To identify the site where the bulk of the energy is pumped into relativistic particles we need to find a signature of accelerated nucleons. Such a signature will be the spectrum reflective of neutral-pion decay from collisions of freshly-accelerated nuclei with nearby gas and dust[2]. Depending of the conditions in the medium the neutral-pion decay $\gamma$ rays will be evident as a spectral feature in the range 50 MeV to a few GeV, while above this range their spectral power law index will match the spectral index of the cosmic rays at the site. This range is ideal for studies with GLAST.

## GLAST CAPABILIES & SNR

The GLAST telescope we have proposed is shown in Figure 1. Its capabilities are summarized in Figure 2. At 1 GeV, the effective area, including inefficiencies due to photon conversion and background rejection, will be more than 1 m$^2$, and our single photon angular resolution will be 0.4 degrees. Supernova remnants can be observed well off axis (up to 60 degrees) without loss of angular resolution and the 2.4 sr solid angle allows any individual source to be observed at a 20% duty cycle. Thus, supernova remnants can be observed in the normal sky-scanning mode of GLAST operation. Minimal tails in the point-spread function (the 95% containment angle is typically 2.5 times the 68% containment angle) optimize the capability for mapping the structure of $\gamma$-ray emission. The challenge will be to find sources with angular scales large enough to be mapped with angular resolution sufficient to separate the extended shell emission from that of the central compact source, nominally a pulsar.

The obvious candidates for this search are those SNR for which EGRET reported a finite flux [6]. We also include RXJ1713.7-3946 for which TeV emission has recently been reported [7]. They are listed in Table 1 along with relevant parameters. To predict the signal GLAST might detect for a typical case, we have modeled the $\gamma$ rays from $\gamma$ Cygni taking the total flux as measured by EGRET and dividing it into two separate components with 60% of the flux from the pulsar [8].

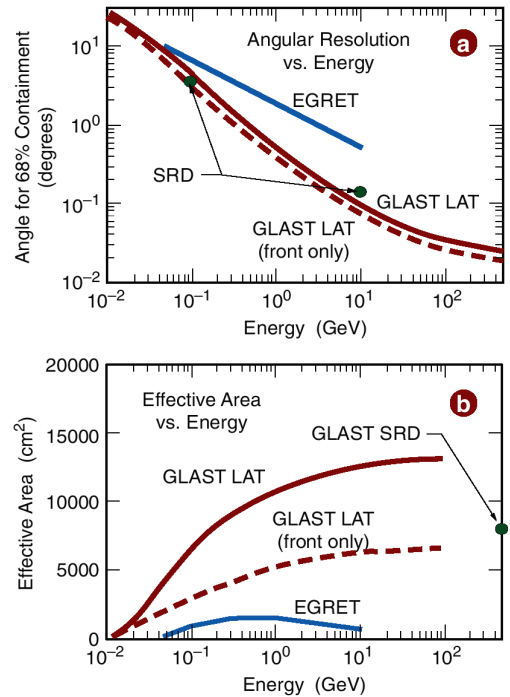

Fig. 2: Performance characteristics of GLAST

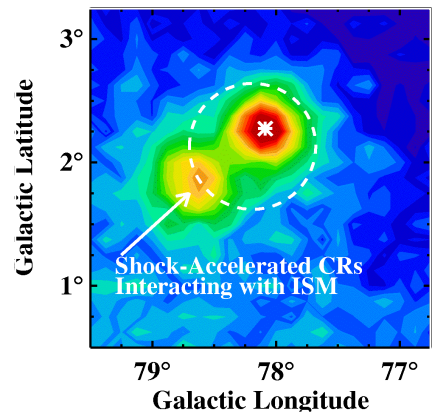

Fig. 3: Simulated observation of $\gamma$ Cygni

| TABLE 1. CANDIDATE SUPERNOVA | | | |
|---|---|---|---|
| Remnant | Distance (kpc) | Age (years) | Size (arcmin) |
| RXJ1713.7-3946 | 6 | ?? | 70 |
| $\gamma$ Cygni | 1.8 | 7000 | 60 |
| IC443 | 1.5 | 5000 | 45 |
| SN1006 | ~3 | 1000 | 30 |
| CasA | 2.8 | 300 | 5 |
| Kepler | 4.4 | 400 | 3 |

---

[2] P+P->$\pi^o$+X and $\pi^o \Rightarrow 2\gamma$

The result of this simulation is shown in Figure 3. Note that the shell component is clearly resolved from the central point source. The spectrum of the γ Cygni region is taken from Ref. 6.

Assuming this source (γ Cygni) is as modeled we expect to spatially resolve the shell source and to be able to measure the spectrum [9] shown in Figure 4. The locally observed cosmic-ray nuclei have spectra proportional to $E^{-2.7}$. This spectrum is the natural result of shock models which predict harder source spectra, $E^{-\delta}$ with spectral exponents δ in the range 2.0 - 2.3 combined with the energy dependent losses by diffusion. The IC and bremsstrahlung components can be deconvolved and subtracted to determine the $\pi^o$ contribution. The asymptotic spectral exponent at energies above 10 GeV should be indicative of the source spectrum of freshly-accelerated cosmic rays.

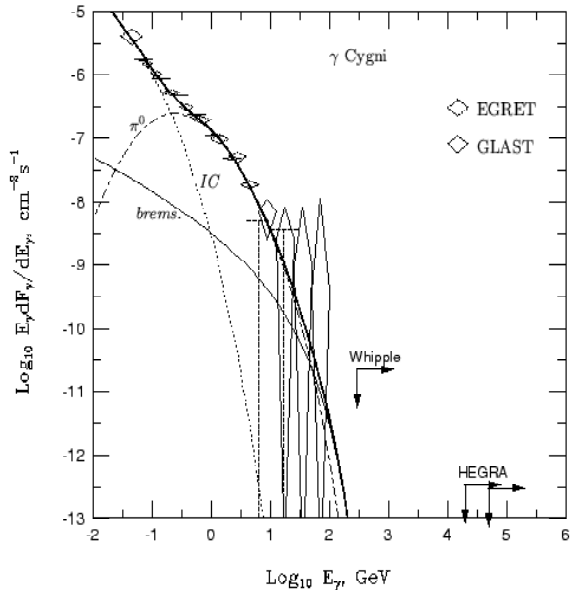

Fig. 4: Expected data from GLAST compared to one of the models of Gaisser et al. [9].

## DIFFUSE CONTINUUM GAMMA-RAYS

Once accelerated, the cosmic rays quickly diffuse away from their sources, mix with those from other sources and diffuse out of the Galaxy. Most theoretical treatments of cosmic-ray transport from their sources to the Earth assume the result of this process is a steady flux of cosmic rays throughout the Galaxy with sources in equilibrium with losses.

With GLAST we have a chance to probe the spatial structure on a scale that can begin to test this hypothesis.

Of the many possible observables of cosmic-ray transport through the interstellar medium to earth (e.g. Be, B isotopes, positrons, antiprotons) the least exploited to date has been γ rays. The diffuse γ-ray emission in any direction is a measure of the point by point product of the cosmic-ray intensity and the matter density in the interstellar medium. The secondary antiprotons and positrons are produced in the same collisional processes. The EGRET team has modeled this diffuse γ-ray emission [10] on 0.5 degree scale. Their models distribute the cosmic rays proportionally to the matter in the Galaxy on scales of order 1 kpc.

One of the most puzzling findings from EGRET is the excess in the Galactic diffuse γ-ray emission above 1 GeV [11]. Independent confirmation and extension toward the higher energies would help to understand this excess. Is the spectrum of cosmic-ray nuclei [12] or electrons [13] elsewhere in the Galaxy different from what we measure locally? If so, what are the corresponding consequences for all cosmic-ray physics? The range of such possibilities have been recently discussed in the literature [14].

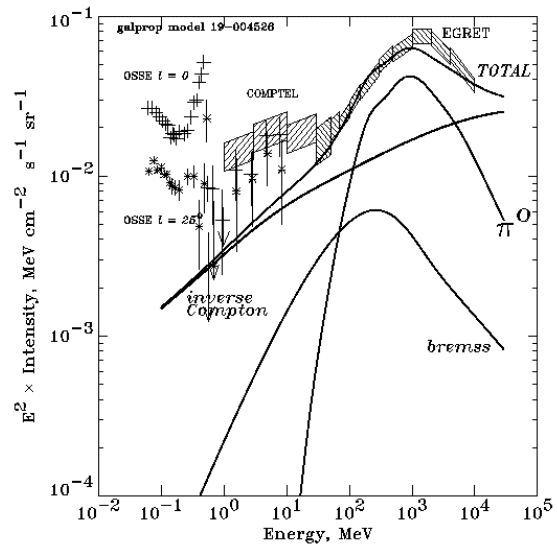

Fig. 5: A model [14] of the Galactic diffuse emission that fits the high energy EGRET data.

The self-consistent approach to this problem developed by Strong et al. [14] models the diffusive propagation of nucleons and electrons. Figure 5 shows their predictions of the various contributing components. Their conclusion, reflected in the figure, is that the harder electron spectrum in the Galaxy is more likely. But their solution is not unique and requires confirmation with the additional data that GLAST will provide. Measurements of high-energy antiproton and positron spectra provide a useful constraint on the nucleon spectrum, while the electron spectrum can be different from place to place due to the large electron energy losses (though radio measurements of synchrotron emission provide a constraint for some electron energies). These hypotheses can be tested with GLAST by measuring the spectrum of Galactic diffuse γ-ray emission above few GeV. First, one can distinguish between the inverse Compton and neutral-pion decay components, and, second, the latitude and longitude γ-ray profiles are different for gas-related nucleon and broader electron components.

Another result of the studies [15] is that the inverse Compton (IC) scattering, especially at high latitudes, plays a more important role than previously thought. The anisotropic distribution of photons above the Galactic plane will result in greater (up to 40%) flux from IC in the halo. Evidence for a large γ-ray halo has been found in an analysis of the EGRET data [16]. This affects our estimates of the extragalactic component, the intensity of cosmic rays in the Galaxy at large and the size of the Galactic cosmic-ray containment halo. The importance of high latitude IC can be tested with GLAST by measuring γ-ray profiles at high energies. Therefore, a major objective of GLAST will be to resolve the high latitude diffuse emission [17] and find, or at least greatly reduce, the fraction of that flux which is due to unresolved point sources.

A longstanding question in cosmic-ray astrophysics has been the extent to which cosmic rays can penetrate into clouds and interact with material there. The issue is one of the length scale of equipartition of energy between cosmic rays, thermal motion of interstellar gas and energy in magnetic fields. Equipartition seems to hold on the kpc scale, but what about scales below 1 kpc? GLAST will be able to study the γ-ray emission from molecular clouds, super-bubble walls and other interstellar concentrations of matter on an angular scale heretofore impossible reaching the tens of arc minute range. At a distance of 1 kpc, spatial scales of 3-10 parsec will be resolved.

## CONCLUSION

Gamma rays are a powerful tool to explore the sources and distribution of cosmic rays in our Galaxy. In turn, having improved understanding of the role of cosmic rays is essential for the study of many topics in γ-ray astronomy to be addressed by GLAST. The results of the studies discussed here will have important consequences for understanding of the distribution of cosmic rays in the Galaxy and their dynamic effects thereupon. Besides, it is worth noting that complete understanding of both the Galactic cosmic rays and the γ-ray emissions are essential for the study of the dark matter in the Galaxy.